%Paper: gr-qc/9406007
%From: Neil Cornish <cornish@medb.physics.utoronto.ca>
%Date: Fri, 3 Jun 1994 19:28:04 -0400

\magnification=1200
\voffset=0 true mm
\hoffset=0 true in
\hsize=6.5 true in
\vsize=8.5 true in
\normalbaselineskip=13pt
\def\doublespace{\baselineskip=20pt plus 3pt\message{double space}}
\def\singlespace{\baselineskip=13pt\message{single space}}
\let\spacing=\singlespace
\parindent=1.0 true cm

% bold face mathe italic fonts in dir 2160, 1800, 1643, and 1500
 %ambi in VAX

% also available in dir 1000,1095,1200,1315, and 1440

\newcount\equationumber \newcount\sectionumber
\sectionumber=1 \equationumber=1
\def\setsection{\global\advance\sectionumber by1 \equationumber=1}

\def\numbe{{{\number\sectionumber}{.}\number\equationumber}
                            \global\advance\equationumber by1}
\def\numberit{\eqno{(\number\equationumber)} \global\advance\equationumber by1}

\def\numberal{(\number\equationumber)\global\advance\equationumber by1}

\def\sectionit{\eqno{(\numbe)}}

\def\ccf#1{\,\vcenter{\normalbaselines
    \ialign{\hfil$##$\hfil&&$\>\hfil ##$\hfil\crcr
      \mathstrut\crcr\noalign{\kern-\baselineskip}
      #1\crcr\mathstrut\crcr\noalign{\kern-\baselineskip}}}\,}
\def\scf#1{\,\vcenter{\baselineskip=9pt
    \ialign{\hfil$##$\hfil&&$\>\hfil ##$\hfil\crcr
      \vphantom(\crcr\noalign{\kern-\baselineskip}
      #1\crcr\mathstrut\crcr\noalign{\kern-\baselineskip}}}\,}

\def\small3j#1#2#3#4#5#6{\def\st{\scriptstyle} % 3j-symbol - small size
   \bigl(\scf{\st#1&\st#2&\st#3\cr
           \st#4&\st#5&\st#6\cr} \bigr)}

   %Name of a nucleus

%\def\slashA{\hbox{$A\mkern-9mu/\mkern 9mu$}}

%\def\slasshA{\hbox{$A\mkern-9mu/\mkern 5mu$}}

\def\ref#1{$^{#1)}$}

   %Figure caption
              %#4 for caption

%...... subscripts and supscripts .....................................
\def\upa#1{\raise 1pt\hbox{\sevenrm #1}}
\def\dna#1{\lower 1pt\hbox{\sevenrm #1}}
\def\dnb#1{\lower 2pt\hbox{$\scriptstyle #1$}}
\def\dnc#1{\lower 3pt\hbox{$\scriptstyle #1$}}
\def\upb#1{\raise 2pt\hbox{$\scriptstyle #1$}}
\def\upc#1{\raise 3pt\hbox{$\scriptstyle #1$}}
\def\hprime{\raise 2pt\hbox{$\scriptstyle \prime$}}
\def\ccom{\,\raise2pt\hbox{,}}

%.... special maths symbols

\def\asymptotically#1{\;\rlap{\lower 4pt\hbox to 2.0 true cm{
    \hfil\sevenrm  #1 \hfil}}
   \hbox{$\relbar\joinrel\relbar\joinrel\relbar\joinrel
     \relbar\joinrel\relbar\joinrel\longrightarrow\;$}}
\def\Asymptotically#1{\;\rlap{\lower 4pt\hbox to 3.0 true cm{
    \hfil\sevenrm  #1 \hfil}}
   \hbox{$\relbar\joinrel\relbar\joinrel\relbar\joinrel\relbar\joinrel
     \relbar\joinrel\relbar\joinrel\relbar\joinrel\relbar\joinrel
     \relbar\joinrel\relbar\joinrel\longrightarrow$\;}}

\catcode`@=11
\def\C@ncel#1#2{\ooalign{$\hfil#1\mkern2mu/\hfil$\crcr$#1#2$}}
\def\gf#1{\mathrel{\mathpalette\c@ncel#1}}      % slash a small letter
\def\Gf#1{\mathrel{\mathpalette\C@ncel#1}}      % slash a big letter

\def\gapx{\lower 2pt \hbox{$\buildrel>\over{\scriptstyle{\sim}}$}}
\def\lapx{\lower 2pt \hbox{$\buildrel<\over{\scriptstyle{\sim}}$}}

\def\nablaleft{\hbox{\raise 6pt\rlap{{\kern-1pt$\leftarrow$}}{$\nabla$}}}
\def\nablaright{\hbox{\raise 6pt\rlap{{\kern-1pt$\rightarrow$}}{$\nabla$}}}
\def\nablaboth{\hbox{\raise 6pt\rlap{{\kern-1pt$\leftrightarrow$}}{$\nabla$}}}

\def\boks#1#2{{\hsize=#1 true cm\parindent=0pt
  {\vbox{\hrule height1pt \hbox
    {\vrule width1pt \kern3pt\raise 3pt\vbox{\kern3pt{#2}}\kern3pt
    \vrule width1pt}\hrule height1pt}}}}

\def\heading{ }
\def\range{ }

\def\body{\vfill\eject\parindent=1.0 true cm
 \ifx\spacing\singlespace\singlespace\else\doublespace\fi}
\def\title#1{\centerline{{\bf #1}}}

\def\today{\ifcase\month\or
  January\or February\or March\or April\or May\or June\or
  July\or August\or September\or October\or November\or December\fi
  \space\number\day, \number\year}
\let\date=\today
\newcount\hour \newcount\minute
\countdef\hour=56
\countdef\minute=57
\hour=\time
  \divide\hour by 60
  \minute=\time
  \count58=\hour
  \multiply\count58 by 60
  \advance\minute by -\count58

\def\sectionskip{\penalty-500\vskip24pt plus12pt minus6pt}

\def\sec{\hbox{\lower 1pt\rlap{{\sixrm S}}{\hbox{\raise 1pt\hbox{\sixrm S}}}}}
\def\section#1\par{\goodbreak\message{#1}
    \sectionskip\nobreak\noindent{\bf #1}\vskip0.3cm \noindent}

\nopagenumbers
\headline={\ifnum\pageno=\count31\frontheadline
  \else{\ifnum\pageno=0\frontheadline
     \else{{\raise 2pt\hbox to \hsize{\paperhead}}}\fi}\fi}
%\headline={\ifnum\pageno=\count31\frontheadline
%  \else{\ifnum\pageno=0\frontheadline
%     \else{\underbar{\raise 2pt\hbox to \hsize{\paperhead}}}\fi}\fi}

\footline={\centerline{\sevenbf \folio}}
\def\frontheadline{\sevenbf \hfil}
\def\paperhead{\sevenbf \heading\ \range\hfil\folio}
\newdimen\pagewidth \newdimen\pageheight \newdimen\ruleht
\maxdepth=2.2pt
\pagewidth=\hsize \pageheight=\vsize \ruleht=.5pt

\def\onepageout#1{\shipout\vbox{ % here we define one page of output
    \offinterlineskip % butt the boxes together
  \makeheadline
    \vbox to \pageheight{
         #1 % now insert the main information
 \ifnum\pageno=\count31{\vskip 21pt\line{\the\footline}}\fi
 \ifvoid\footins\else %footnot ino is present
 \vskip\skip\footins \kern-3pt
 \hrule height\ruleht width\pagewidth \kern-\ruleht \kern3pt
 \unvbox\footins\fi
 \boxmaxdepth=\maxdepth}
 \advancepageno}}
\output{\onepageout{\pagecontents}}
\count31=-1
\topskip 0.7 true cm
\pageno=0
\doublespace
\centerline{\bf Non-Singular Gravity Without Black Holes}
\centerline{\bf N. J. Cornish and J. W. Moffat}
\centerline{\bf Department of Physics}
\centerline{\bf University of Toronto}
\centerline{\bf Toronto, Ontario M5S 1A7}
\centerline{\bf Canada} \vskip 1 true in
\centerline{``What are the everywhere regular solutions of these field
equations?''}
\centerline{A. Einstein, Autobiographical notes, 1949.}
\vskip 1 true in
\centerline{May, 1994.}
\vskip 2 true in
{\bf UTPT-94-08}
\vskip 0.1 true in
{\bf e-mail: Moffat@medb.physics.utoronto.ca}
\vskip 0.1 true in
\hskip 0.65 true in {\bf Cornish@medb.physics.utoronto.ca}
\par\vfil\eject
\centerline{\bf Non-Singular Gravity Without Black Holes}
\centerline{\bf N. J. Cornish and J. W. Moffat}
\centerline{\bf Department of Physics}
\centerline{\bf University of Toronto}
\centerline{\bf Toronto, Ontario M5S 1A7}
\centerline{\bf Canada}
\vskip 0.4 true in
\centerline{\bf Abstract}
\vskip 0.2 true in
A non-singular static spherically symmetric solution of the nonsymmetric
gravitational and electromagnetic theory (NGET)
field equations is derived, which depends on the four parameters $m, \ell^2,
Q$ and $s$, where $m$ is the mass, $Q$ is the electric charge, $\ell^2$ is
the NGT charge of a body and $s$ is a dimensionless constant.  The
electromagnetic field invariant is also singularity-free, so that it is
possible to construct regular particle-like solutions in the theory.
All the curvature invariants are finite, there are no null surfaces in the
spacetime and there are no black holes. A new stable, superdense object
(SDO) replaces black holes.
\par\vfil\eject
\proclaim 1. {\bf Introduction} \par
\vskip 0.2 true in
After publishing his theory of gravitation in 1916$^{1}$, Einstein set himself
the goal of finding a unified field theory of electromagnetism and gravitation.
However, a more pressing issue for him was the discovery of everywhere
regular solutions of such a unified field theory$^{2}$. He failed to discover
a satisfactory unification of gravitation and electromagnetism but in his
search he developed a unified field theory based on a
nonsymmetric field structure$^{3}$. In 1979, it was proposed by one of
us$^{4-6}$ that the nonsymmetric field structure had nothing to do with
electromagnetism but instead was a general description of the geometry
of spacetime, i.e. it describes a theory of the {\it pure} gravitational
field called the nonsymmetric gravitational theory (NGT).

Although Einstein's theory has proved to be in good agreement with all the
experimental tests that it has been subjected to so far, there has always
been the issue of the existence of singularities in the solutions of the field
equations. There exist null surfaces or event horizons in the
solutions at which the red-shift becomes infinite. Although physics has learned
to live with these event horizons, and the notion of cosmic censorship was
invented to prevent an observer at infinity from seeing a ``naked''
singularity,
nobody has succeeded in proving rigorously that such a cosmic censorship
exists, and naked singularity solutions of Einstein's field equations have
been published$^{7}$. Such a naked singularity would destroy the Cauchy
data on an initial space-like surface, and due to the local nature of the
solution for gravitational collapse would invalidate Einstein's theory.
The big-bang singularity in the cosmological solutions of the theory
have also been cause for concern among theorists, although due to the
global nature of these solutions such a singularity would not vitiate the
theory to the same extent as the gravitational collapse solution.

There has been much discussion recently about paradoxes that occur in
connection with Hawking radiation and event horizons. An observer at spatial
infinity would see infalling matter ``freeze'' before it can form an event
horizon, whereas a freely falling observer can fall through an event horizon
without difficulty, but be unable to communicate this fact to the observer
at infinity. This means that spacetime is separated into two disconnected
parts by a null surface and there exists no communication between the
two spacetimes. A paradox arises, for these two observers completely
disagree about what they see in a spacetime containing a black hole.
The Hawking radiation is purely thermal and contains no
information about a collapsed star. Thus, as first pointed out by
Hawking$^{8}$, this leads to what has been called the information loss
paradox. Neither of these paradoxes has been resolved satisfactorily
in spite of attempts to do so by modifying quantum mechanics and invoking
an, as yet, unknown theory of quantum gravity.

An alternative way to avoid these paradoxes is to modify Einstein gravity
theory (EGT), so that black holes and singularities are eliminated altogether.
This cannot be done in a gravity theory which contains new degrees of
freedom arising from some new conserved charge for which there exists a
{\it smooth} limit to GR as the charge tends to zero. An attempt to avoid
black hole solutions in NGT, using only solutions depending on the $\ell^2$
charge failed for this reason$^{9}$, for photons do not have an $\ell^2$
charge and, therefore, a pure photon star is described by GR and can collapse
to a black hole. Moreover, there is no reason why in every given situation
the parameter $\ell^2$ for a star should have a value that forbids collapse
to a black hole. The aforementioned NGT solutions were based on a simplifying
assumption about the form of the skew symmetric fields. It is only when this
assumption is discarded that the true non-singular nature of NGT is revealed.

In the following, the complete non-singular solution of the NGT field
equations is derived, which for non-vanishing values of a dimensionless
parameter $s$ has no null surfaces in spacetime and the curvature invariants
are finite$^{10}$. For strong fields, the dependence on the parameter $s$ is
non-analytic, so that
a smooth limit to Einstein's gravitational theory does not exist. This
means that there is a true absence of black holes in NGT, since even for
infinitesimally small $s$ there are no event horizons in the spacetime
and no singularity at $r=0$.
Einstein's gravity theory is a separate theory with singularities, which
exists only when it is assumed that $g_{\mu\nu}$ and $\Gamma^\lambda_{\mu\nu}$
are symmetric in the indices $\mu$ and $\nu$. For sufficiently small $s$ and
$\ell^2$ charge, NGT agrees with all current experiments.

There are no black holes, wormholes or other exotic objects in the nonsingular
NGT, and the stable superdense objects (SDO's)$^{11}$ that replace black holes
for $r\leq 2m$ will not exhibit infinite red-shifts at their surfaces.
Therefore, all observers will agree on what happens when a star collapses to a
SDO, and no matter can disappear into a singularity in the interior of a
SDO. There will not be any Hawking radiation from the surface of a SDO;
only normal radiation of both a thermal and non-thermal nature will be
emitted from the surface. Stable neutron stars exist for
$s\leq 15$$\,$$^{11}$. Larger values of $s$ will cause neutron stars to be
unstable
against gravitational {\it expansion}, not collapse. When the NGT $\ell^2$
charge effects are included in the neutron star calculations$^{12}$, then
it is possible to counterbalance the attractive effects of $\ell^2$ against
the repulsive effects of $s$, weakening the bounds on the coupling of NGT
charge
to matter, and the bound on $s$.

In Section 2., we shall present the Lagrangian density and the field
equations of the nonsymmetric gravitational, electromagnetic theory (NGET).
The general properties of the nonsymmetric static spherically symmetric
solutions will be discussed in Section 3, while in Section 4 we analyze the
non-singular nature of these solutions. We conclude with a summary of the
results in Section 5.
\par\vfil\eject
\setsection\proclaim 2.
{\bf NGT Lagrangian Density and Field Equations}
\par \vskip 0.2 true in
The Lagrangian density including electromagnetism and sources, in NGT, is given
by$^{4,6,13,14}$:
$$
{\cal L} = {\bf g}^{\mu\nu} R_{\mu\nu} (W) + \sqrt{-g}[\kappa(g^{[\mu\nu]}
F_{\mu\nu})^2-H^{\mu\nu}F_{\mu\nu}]+{\cal L}_M,
\sectionit
$$
where ${\bf g}^{\mu\nu}=\sqrt{-g}g^{\mu\nu}$ and
$R_{\mu\nu}(W)$ is the NGT contracted curvature tensor:
$$
R_{\mu\nu}(W)=W^\beta_{\mu\nu,\beta} - {1\over
2}(W^\beta_{\mu\beta,\nu}+W^\beta_{\nu\beta,\mu}) -
W^\beta_{\alpha\nu}W^\alpha_{\mu\beta} +
W^\beta_{\alpha\beta}W^\alpha_{\mu\nu},
\sectionit
$$
defined in terms of the unconstrained nonsymmetric connection:
$$
W^\lambda_{\mu\nu}=\Gamma^\lambda_{\mu\nu}-{2\over 3}\delta^\lambda_\mu
W_\nu,
\sectionit
$$
where $W_\mu\equiv W^\lambda_{[\mu\lambda]} = {1\over 2}
(W^\lambda_{\mu\lambda}-W^\lambda_{\lambda\mu})$.  This equation
leads to:
$$
\Gamma_\mu=\Gamma^\lambda_{[\mu\lambda]}=0.
\sectionit
$$
The skew tensor $H_{\mu\nu}=-H_{\nu\mu}$ is defined in terms of the skew
electromagnetic field tensor $F_{\mu\nu}$ by the equation:
$$
g_{\sigma\beta}g^{\gamma\sigma}H_{\gamma\alpha}+g_{\alpha\sigma}
g^{\sigma\gamma}H_{\beta\gamma}=2g_{\alpha\sigma}g^{\sigma\gamma}F_{\beta
\gamma}
\sectionit
$$
and $\kappa$ is a coupling constant.
The contravariant tensor $g^{\mu\nu}$ is defined in terms of the
equation:
$$
g^{\mu\nu}g_{\sigma\nu}=g^{\nu\mu}g_{\nu\sigma}=\delta^\mu_\sigma.
\sectionit
$$
The NGT contracted curvature tensor can be written as
$$
R_{\mu\nu}(W) = R_{\mu\nu}(\Gamma) + {2\over 3}
W_{[\mu,\nu]},
\sectionit
$$
where $R_{\mu\nu}(\Gamma)$ is defined by
$$
R_{\mu\nu}(\Gamma ) = \Gamma^\beta_{\mu\nu,\beta} -{1\over
2} \left(\Gamma^\beta_{(\mu\beta),\nu} +
\Gamma^\beta_{(\nu\beta),\mu}\right) - \Gamma^\beta_{\alpha\nu}
\Gamma^\alpha_{\mu\beta} +
\Gamma^\beta_{(\alpha\beta)}\Gamma^\alpha_{\mu\nu}.
\sectionit
$$

The Lagrangian density for the matter sources is given by (G=c=1):
$$
{\cal L}_M= -8\pi g^{\mu\nu} {\bf T}_{\mu\nu} + {8\pi \over 3}
W_\mu {\bf S}^\mu.
\sectionit
$$
When only electromagnetic fields and gravitation exist in the vacuum i.e.
in the absence of phenomenological matter sources, then $S^\mu=0$ and the
electromagnetic energy-momentum tensor is given by$^{4,13,14}$:
$$
T_{\alpha\beta}=-{1\over 4\pi}[(g_{\sigma\beta}H^{\mu\sigma}F_{\mu\alpha}
-\kappa g^{[\mu\nu]}F_{\mu\nu}F_{\alpha\beta}-{1\over 4}g_{\alpha\beta}
(H^{\mu\nu}H_{\mu\nu}-\kappa(g^{\mu\nu}F_{\mu\nu})^2)],
\sectionit
$$
where
$$
H^{\mu\alpha}=g^{\beta\mu}g^{\gamma\alpha}H_{\beta\gamma}.
\sectionit
$$
It can be proved that
$$
g^{\alpha\beta}T_{\alpha\beta}=0.
\sectionit
$$
We observe that there is a coupling term of the form $\kappa
g^{[\mu\nu]}F_{\mu\nu}$ in the Lagrangian density. However, the non-singular
nature of the theory is manifest for any $\kappa$, including $\kappa=0$.

Our field equations are given by
$$
G_{\mu\nu} (W) = 8\pi T_{\mu\nu},
\sectionit
$$
$$
{{\bf g}^{[\mu\nu]}}_{,\nu} = 0,
\sectionit
$$
$$
g_{\mu\nu,\sigma}-g_{\rho\nu}\Gamma^\rho_{\mu\sigma}-
g_{\mu\rho}\Gamma^\rho_{\sigma\nu}=0,
\sectionit
$$
$$
({\bf H}^{\alpha\mu}-\kappa{\bf g}^{[\alpha\mu]}g^{\nu\beta}
F_{\nu\beta})_{,\mu}=0,
\sectionit
$$
where
$$
G_{\mu\nu} = R_{\mu\nu} - {1\over 2} g_{\mu\nu}R.
\sectionit
$$

The variational principle yields for invariance under coordinate
transformations the four Bianchi identities:
$$
\left[{\bf g}^{\alpha\nu} G_{\rho\nu}(\Gamma) + {\bf g}^{\nu\alpha}
G_{\nu\rho} (\Gamma)\right]_{,\alpha} + {g^{\mu\nu}}_{, \rho} {\bf
G}_{\mu\nu}(\Gamma) = 0.
\enskip
\sectionit
$$
The matter response equations are
$$
{1\over 2}\left(g_{\sigma\rho}{\bf T}^{\sigma\alpha} +
g_{\rho\sigma}{\bf T}^{\alpha\sigma}\right)_{,\alpha} - {1\over 2}
g_{\alpha\beta,\rho}{\bf T}^{\alpha\beta} = 0.
\sectionit
$$

For ${\bf S}^\mu$ nonzero, Eq.(2.14) becomes
$$
{{\bf g}^{[\mu\nu]}}_{,\nu}=4\pi {\bf S}^\mu.
\sectionit
$$
If we perform a Hodge decomposition of ${\bf g}^{[\mu\nu]}$:
$$
{\bf g}^{[\mu\nu]}={\bf a}^{[\mu,\nu]} + \epsilon^{\mu\nu\kappa\lambda}
{\bf b}_{[\kappa,\lambda]},
\sectionit
$$
we find from (2.20) that the three degrees of freedom  ${\bf a}^\mu$
are determined by the NGT charge current ${\bf S}^\mu$. The other three
components, ${\bf b}^\mu$, of ${\bf g}^{[\mu\nu]}$ are not directly coupled
to the NGT charge. It is the degrees of freedom associated with
${\bf b}^\mu$ which give rise to the non-singular nature of NGT (refered to
as NSG in Refs.[10,11]). Previous work on NGT concentrated on the
${\bf a}^\mu$ degrees of freedom.
\vskip 0.2 true in
\setsection\proclaim 3. {\bf The Static Spherically Symmetric Solutions}
\par
\vskip 0.2 true in
In the case of a static spherically symmetric field,
Papapetrou has derived the canonical form of $g_{\mu\nu}$$^{15}$:
$$
g_{\mu\nu}=\left(\matrix{-\alpha&0&0&w\cr
0&-\beta&f\hbox{sin}\theta&0\cr 0&-f\hbox{sin}\theta&
-\beta\hbox{sin}^2
\theta&0\cr-w&0&0&\gamma\cr}\right),
\sectionit
$$
where $\alpha,\beta,\gamma$ and $w$ are functions of $r$. The
tensor $g^{\mu\nu}$ has the components:
$$
g^{\mu\nu}=\left(\matrix{{\gamma\over w^2-
\alpha\gamma}&0&0&{w\over w^2-\alpha\gamma}\cr
0&-{\beta\over \beta^2+f^2}&{f\hbox{csc}\theta\over \beta^2+f^2}&0\cr
0&-{f\hbox{csc}\theta\over \beta^2+f^2}&-{\beta\hbox{csc}^2\theta\over
\beta^2+f^2}&0\cr-{w\over w^2-\alpha\gamma}&0&0&-{\alpha\over
w^2-\alpha\gamma}\cr}\right).
\sectionit
$$

The electromagnetic field $F_{\mu\nu}$ is defined in terms of the
potentials $A_\mu$:
$$
F_{\mu\nu}=A_{\nu,\mu}-A_{\mu,\nu},
\sectionit
$$
and it has the static components:
$$
F_{10}=E(r),\quad F_{23}=H(r)\,\hbox{sin}\theta,
\sectionit
$$
all other components being zero. From (2.5) and (3.1)-(3.4), it follows that
$H_{\mu\nu}=F_{\mu\nu}$ and from the equation:
$$
F_{\mu\nu,\sigma}+F_{\nu\sigma,\mu}+F_{\sigma\mu,\nu}=0,
\sectionit
$$
we find that $H(r)$ is a constant, which corresponds to the magnetic charge.
We shall assume in accordance with Maxwell's theory that the magnetic charge
is zero. We have
$$
H^{10}=-{E\over \alpha\gamma-w^2}.
\sectionit
$$
The determinant of the $g_{\mu\nu}$ is given by
$$
\sqrt{-g}=\hbox{sin}\theta(\alpha\gamma-w^2)^{1/2}(\beta^2+f^2)^{1/2}.
\sectionit
$$

The solution to Eq. (2.14) is
$$
w^2={\ell^4\alpha\gamma\over \beta^2+f^2+\ell^4},
\sectionit
$$
where $\ell^2$ is a constant of integration which is identified with the
NGT charge. Eq. (2.16) has the solution:
$$
E=\biggl({w\over \ell^2}\biggr)\biggl({Q\rho^2\over \rho^2+\kappa^2\ell^4}
\biggr)={Q\rho\sqrt{\alpha\gamma-w^2}\over
(\rho^2+\kappa^2\ell^4)},
\sectionit
$$
where $Q$ is the electric charge of a particle and
$$
\rho^2=\beta^2+f^2.
\sectionit
$$

For $\ell^2=0$, we have
$$
E={Q\sqrt{\alpha\gamma}\over \rho}.
\sectionit
$$

The field equations (2.13) for the static spherically symmetric case take the
form$^{12}$:
$$
1+\biggl({fB^\prime-\beta A^\prime\over 2\alpha}\biggr)^\prime
+B^\prime\biggl({\beta B^\prime+fA^\prime\over 2\alpha}\biggr)+
{1\over 2}\biggl({fB^\prime-\beta A^\prime\over 2\alpha}\biggr)
\hbox{ln}(\alpha\gamma U)^\prime=
$$
$$
{\beta\over \rho^2}\biggl({E\ell^2\over w}\biggr)^2+
2\kappa\beta\biggl({Q\ell^2\over \rho^2+\kappa^2\ell^4}\biggr)^2,
\sectionit
$$
$$
c+\biggl({\beta B^\prime+fA^\prime\over 2\alpha}\biggr)^\prime
-B^\prime\biggl({fB^\prime-\beta A^\prime\over 2\alpha}\biggr)+
{1\over 2}\biggl({\beta B^\prime+fA^\prime\over 2\alpha}\biggr)\hbox{ln}
(\alpha\gamma U)^\prime
$$
$$
=-{f\over \rho^2}\biggl({E\ell^2\over w}\biggr)^2-
2\kappa f\biggl({Q\ell^2\over \rho^2+\kappa^2\ell^4}\biggr)^2,
\sectionit
$$
$$
-A^{''}+{1\over 2}(\hbox{ln}\alpha)^\prime A^\prime-{1\over 2}[(A^\prime)^2
+(B^\prime)^2]-{1\over 2}\hbox{ln}(\gamma U)^{''}
+{1\over 4}\hbox{ln}(\gamma U)^\prime \hbox{ln}({\alpha\over \gamma U})^\prime
$$
$$
=-{\alpha\over \rho^2}\biggl({E\ell^2\over w}\biggr)^2+2\kappa\alpha
\biggl({Q\ell^2\over \rho^2+\kappa^2\ell^2}\biggr)^2,
\sectionit
$$
$$
{\gamma \over 2\alpha} [(1-U)[(A^\prime)^2+(B^\prime)^2]
+{1\over 2}(\hbox{ln}\gamma)^\prime\hbox{ln}\biggl({\gamma\rho^2\over
\alpha}\biggr)^\prime
$$
$$
{1\over 2}(\hbox{ln}U)^\prime\hbox{ln}\biggl({\gamma\rho^4\over
\alpha^2}\biggr)^\prime
+{1\over 2}{2U-1\over 1-U}
[(\hbox{ln}U)^\prime]^2+\hbox{ln}(\gamma U^2)^{''}]
$$
$$
={\gamma\over \rho^2}\biggl({E\ell^2\over w}\biggr)^2
-2\kappa\gamma\biggl({Q\ell^2\over \rho^2+\kappa^2\ell^4}\biggr)^2.
\sectionit
$$
Here, $A,B$ and $U$ are defined by
$$
A=\hbox{ln}\rho,\quad B=\hbox{tan}^{-1}\biggl({\beta\over f}\biggr),
\sectionit
$$
$$
U={\rho^2\over \ell^4+\rho^2}=1-{w^2\over \alpha\gamma},
\sectionit
$$
and $A^\prime=\partial A/\partial r$.

It is convenient to use the notation$^{15-18}$:
$$
x={\rho^2\over \alpha},\quad y=\gamma U,\quad \hbox{exp}(q)
=\hbox{exp}(A+iB)=f+i\beta.
\sectionit
$$
Then the field equations can be written in the form:
$$
2A^{''}-(A^\prime)^2+(B^\prime)^2+A^\prime \hbox{ln}(x/y)=0,
\sectionit
$$
$$
(\hbox{ln}y)^{''}+{1\over 2}(\hbox{ln}y)^\prime\hbox{ln}(xy)^\prime
={2\over x}F,
\sectionit
$$
$$
q^{''}+{1\over 2}q^\prime\hbox{ln}(xy)^\prime+2(i+c){\hbox{exp}(q)\over x}
=\biggl({\hbox{exp}(q)\over x}\biggr)G-{2\over x}F,
\sectionit
$$
where
$$
G=-8\kappa\hbox{exp}(-q)\biggl[{Q\ell^2\over \rho^2+\kappa^2\ell^4}\biggr]^2
\sectionit
$$
$$
F=\biggl({E\ell^2\over w}\biggr)^2
-2\kappa\rho^2\biggl[{Q\ell^2\over \rho^2+\kappa^2\ell^4}\biggr]^2.
\sectionit
$$

Let us define:
$$
\lambda(r)=(y^\prime)^2\biggl({x\over y}\biggr).
\sectionit
$$
Then, Eq.(3.21) can be written as
$$
2{d^2p\over dz^2}\lambda+{d\lambda\over dz}{dp\over dz}
+4(i+c)\hbox{exp}(p)=2G\hbox{exp}(p),
\sectionit
$$
where $q+z=p$, $z=\hbox{ln}y$ and $c$ is a constant. An integral of (3.26)
is given by
$$
\biggl({dp\over dz}\biggr)^2\lambda+4(i+c)\hbox{exp}(p)=
\int 2G{d\hbox{exp}(p)\over dz}dz+c_1,
\sectionit
$$
where $c_1$ is a complex constant. We require that
$$
{d\lambda\over dz}=4\hbox{exp}(z)F
=2\hbox{Re}\biggl[\int\hbox{exp}(p){dG\over
dz}\biggr]-(\hbox{Re}\,c_1)+\lambda.
\sectionit
$$

We shall consider the solution for which
$c_1=\lambda_0(1+is)$ where $\lambda_0$ and $s$ are real constants.
We choose as a further boundary condition that
$$
f\rightarrow f_0\quad \hbox{as}\quad  r\rightarrow\infty.
\sectionit
$$
To guarantee that we obtain the Reissner-Nordstrom solution$^{19}$:
$$
\gamma=1-{2m\over r}+{Q^2\over r^2},\quad
\alpha=\biggl(1-{2m\over r}+{Q^2\over r^2}\biggr)^{-1},\quad
\beta=r^2,
\sectionit
$$
when $f=\ell^2=0$, we require that $c=0$.

When $Q=0$, we obtain the Vanstone solution$^{18}$:
$$
f+i\beta=\biggl[{i\lambda_0\over 4y}\biggr](1+is)\hbox{csch}^2
[\sqrt{1+is/2}\hbox{ln}y],
\sectionit
$$
$$
\gamma=\biggl({\ell^4+f^2+\beta^2\over f^2+\beta^2}\biggr)y,\quad
\alpha={(y^\prime)^2(f^2+\beta^2)\over y(4yQ^2+\lambda_0)},
\sectionit
$$
$$
w={\ell^2(y^\prime)\over \sqrt{4yQ^2+\lambda_0}},
\sectionit
$$
where $\lambda_0=4m^2$ and $y$ is an arbitrary function of $r$.

For $Q\not=0$ and $\ell^2=0$, Eq.(3.23) gives
$$
G=0,\quad F=Q^2 = \hbox{const}.
\sectionit
$$
and (3.28) leads to
$$
\lambda=4\hbox{exp}(z),\quad (\hbox{Re}\,c_1)=\lambda_0,
\sectionit
$$
 From (3.26), we get
$$
{2\over \sqrt{c_1}}\hbox{arcsinh}\biggl[\sqrt{ic_1\over 4}
\hbox{exp}(-p/2)\biggr]=\int{dy\over y(4yQ^2+\lambda_0)^{1/2}}.
\sectionit
$$

Solutions to the NGET field equations have been found by Mann$^{14}$,
in the cases
$B^\prime=0$ and $\ell^2=0$. We shall only be concerned here with the latter
solution. We must choose realistic boundary conditions. For $r\rightarrow
\infty$ these must include:
$$
\alpha\rightarrow 1,\quad \gamma\rightarrow 1,\quad \beta\rightarrow r^2.
\sectionit
$$

Choosing $\beta=r^2$ so that the radial coordinate satisfies:
(r-coordinate)=(proper circumference)/$2\pi$, we find that
$$
y=\gamma=\hbox{exp}(\nu),\quad \alpha={(\gamma^\prime)^2(f^2+r^4)\over
\gamma(4\gamma Q^2+\lambda_0)}
\sectionit
$$
$$
f=({\lambda_0\over 2\gamma})(\hbox{cosh}\psi_a-\hbox{cos}\psi_b)^{-2}
[s(1-\hbox{cosh}\psi_a\hbox{cos}\psi_b)+\hbox{sinh}\psi_a\hbox{sin}\psi_b],
\sectionit
$$
where
$$
\psi_a=2a\biggl(\hbox{arcsinh}\sqrt{\lambda_0\over 4Q^2}-
\hbox{arcsinh}\sqrt{\lambda_0\over 4Q^2\gamma}\biggr),\quad \psi_b=
\psi_a(a\leftrightarrow b),
\sectionit
$$
and
$$
a=\sqrt{{\sqrt{1+s^2}+1\over 2}},\quad b=\sqrt{{\sqrt{1+s^2}-1\over 2}}.
\sectionit
$$
Moreover, we have
$$
\lambda_0=4(m^2-Q^2).
\sectionit
$$
The function $\nu$ is given implicitly by
$$
2\hbox{exp}(\nu)(\hbox{cosh}\psi_a-\hbox{cos}\psi_b)^2{r^2\over \lambda_0}=
[s\hbox{sinh}\psi_a\hbox{sin}\psi_b-(1-\hbox{cosh}\psi_a\hbox{cos}\psi_b)].
\sectionit
$$

For $Q=\ell^2=0$, we recover the Wymann solution$^{16}$:
$$
\gamma=\hbox{exp}(\nu),
\sectionit
$$
$$
\alpha=m^2(\nu^\prime)^2\hbox{exp}(-\nu)(1+s^2)\
(\hbox{cosh}(a\nu)-\hbox{cos}(b\nu))^{-2},
\sectionit
$$
$$
f=[2m^2\hbox{exp}(-\nu)(\hbox{sinh}(a\nu)\hbox{sin}(b\nu)+s(1-\hbox{cosh}(a\nu)
\hbox{cos}(b\nu))](\hbox{cosh}(a\nu)-\hbox{cos}(b\nu))^{-2},
\sectionit
$$
where now $\nu$ is implicitly determined by the equation:
$$
\hbox{exp}(\nu)(\hbox{cosh}(a\nu)-\hbox{cos}(b\nu))^2{r^2\over 2m^2}=
\hbox{cosh}(a\nu)\hbox{cos}(b\nu)-1+s\hbox{sinh}(a\nu)\hbox{sin}(b\nu).
\sectionit
$$
\vskip 0.2 true in
\setsection\proclaim 4. {\bf Analysis of the Non-Singular Solutions}
\par
\vskip 0.2 true in
We shall be interested in the branch of multiple solutions for $\nu$ in
Eqs. (3.42) and (3.46), which matches onto the unique solution for large $r$.
Thus, we are interested in the unique inversions of (3.42) and (3.46)
which yield an
asymptotically flat spacetime. We are only able to invert (3.43) analytically
for $r/m < 1, r/Q < 1$ and $2m/r < 1, Q/r < 1$ and we must resort to
numerical methods to establish the intermediate behavior.

We find for $2m/r < 1, Q/r < 1$ and $0 < sm^2/r^2 < 1$ that the metric takes
the
near Reissner-Nordstrom form:
$$
\gamma=1-{2m\over r}+{Q^2\over r^2}+{s^2(m^2-Q^2)^2\over
15r^4}\biggl({m\over r}+{4m^2-Q^2\over r^2}+...\biggr),
\sectionit
$$
$$
\alpha=\biggl[1-{2m\over r}+ {Q^2\over r^2}+{s^2(m^2-Q^2)^2\over 9r^4}
\biggl(2+{7m\over r}+...\biggr)\biggr]^{-1},
\sectionit
$$
$$
f=s(m^2-Q^2)\biggl[{1\over 3}+{2m\over 3r}+{6m^2-Q^2\over 5r^2}+...\biggr],
\sectionit
$$
where the higher order terms in $m/r$ and $Q/r$ include higher powers of
$s$ also. We observe that for small enough $s$ the NGT corrections to
the Reissner-Nordstrom solution and for $Q=0$ to the Schwarzschild
solution, can be made arbitrarily close to experimental predictions of
Einstein-Maxwell theory and EGT.

We can develop expansions near the origin where $r/m < 1, r/Q < 1$ and
$0 < s < 1, Q/m < 1$. Similar expansions exist for $-1 < s <0$. The leading
terms are:
$$
\gamma=\gamma_0+{\cal O}\biggl(\biggl({r\over m}\biggr)^2\biggr)
\sectionit
$$
$$
\alpha={4\over s^2}\biggl(1+{1\over 2}{Q^2\over m^2}\biggr)
\hbox{exp}\biggl(-{\pi\over s}-2-{\pi s\over 8}\biggr)
\biggl({r\over m}\biggr)^2 + {\cal O}\biggl(\biggl({r\over m}\biggr)^4\biggr).
\sectionit
$$
$$
f=m^2\biggl(4-{s\pi\over 2}+s^2- {Q^2\over m^2}\biggr)+
{\cal O}\biggl(\biggl({r\over m}\biggr)^2\biggr),
\sectionit
$$
where $\gamma_0$ is given by
$$
\gamma_0=\biggl(1-{1\over 2}{Q^2\over m^2}\biggr)
\hbox{exp}\biggl(-{\pi\over s}-2 -{\pi s\over 8}\biggr).
\sectionit
$$

For $r$ near zero, $0 < s < 1$ and $(m-Q)$ small, we get
$$
\gamma=\gamma_0+{\cal O}\biggl(\biggl({r\over m}\biggr)^2\biggr),
\sectionit
$$
$$
\alpha={4\gamma_0\over s^2}{r^2\over m^2-Q^2}
+ {\cal O}\biggl(\biggl({r\over m}\biggr)^4\biggr),
\sectionit
$$
$$
f=Q^2\biggl(1-{\pi s\over 8}
+{s^2\over 4}(1-\hbox{ln}2)+...\biggr)+ {\cal O}\biggl(\biggl({r\over
m}\biggr)^2\biggr).
\sectionit
$$
Here, we have
$$
\gamma_0=\biggl({m^2-Q^2\over Q^2}\biggr)\hbox{exp}(-{\pi\over s}-2)
+ {\cal O}\biggl(\biggl({r\over m}\biggr)^2\biggr).
\sectionit
$$

The above results (4.4)--(4.11) clearly illustrate the non-analytic nature
of the
NGET solution in the limit $s\rightarrow 0$ in the limit of strong
gravitational
fields. Numerical results have confirmed that all the expansions given above
produce excellent approximations to the exact solution in their respective
regions of validity.

When $Q=0$, we find for $2m/r < 1$ and $0 < sm^2/r^2 < 1$ that the metric
takes
the near-Schwarzschild form$^{10}$:
$$
\gamma=1-{2m\over r}+{s^2m^5\over 15r^5}+{4s^2m^6\over 15r^6}+...,
\sectionit
$$
$$
\alpha=\biggl(1-{2m\over r}+{2s^2m^4\over 9r^4}+{7s^2m^5\over 9r^5}
+...\biggr)^{-1},
\sectionit
$$
$$
f={sm^2\over 3}+{2sm^3\over 3r}+{6sm^4\over 5r^2}+....
\sectionit
$$
Near $r=0$ we can develop expansions where $r/m < 1$ and $0 < \vert s\vert
<1$.
The leading terms are
$$
\gamma=\gamma_0+{\gamma_0(1+{\cal O}(s^2))\over 2\vert s\vert}\biggl({r\over
m}\biggr)^2 + {\cal O}\biggl(\biggl({r\over m}\biggr)^4\biggr)
\sectionit
$$
$$
\alpha={4\gamma_0(1+{\cal O}(s^2))\over s^2}\biggl({r\over m}\biggr)^2
+{\cal O}\biggl(\biggl({r\over m}\biggr)^4\biggr),
\sectionit
$$
$$
f=m^2\biggl(4-{\vert s\vert\pi\over 2}+s\vert s\vert+{\cal O}(s^3)\biggr)
+{\vert s\vert+s^2\pi/8+{\cal O}(s^3)\over 4}r^2+{\cal O}(r^4),
\sectionit
$$
$$
\gamma_0=\hbox{exp}\biggl(-{\pi+2s\over \vert s\vert}+{\cal O}(s)\biggr)...
\sectionit
$$
As in the electrically charged case, these solutions clearly illustrate
the non-analytic nature of the limit $s\rightarrow 0$ in the strong
gravitational field regime.

The solution for the extremal case $Q=m$ cannot be obtained from the above
expansions, but must be derived from another branch of the non-singular
solution given by:
$$
f={s\over 2\gamma}(1-\hbox{cosh}\psi\hbox{cos}\psi)
(\hbox{cosh}\psi-\hbox{cos}\psi)^{-2},
\sectionit
$$
$$
2\gamma
r^2=s\hbox{sinh}\psi\hbox{sin}\psi(\hbox{cosh}\psi-\hbox{cos}\psi)^{-2},
\sectionit
$$
where
$$
\psi=\sqrt{s\over 2}\biggl({1\over \sqrt{\gamma}}-1\biggr).
\sectionit
$$
Near $r=0$ we get
$$
\gamma={s\over (\sqrt{s}+\pi\sqrt{2})^2}+{\cal O}\biggl(
\biggl({r\over Q}\biggr)^2\biggr),
\sectionit
$$
$$
\alpha={8(\hbox{cosh}\pi+1)\over (\hbox{cosh}\pi-1)(\sqrt{s}+\pi\sqrt{2})^2}
\biggl({r\over Q}\biggr)^2+{\cal O}\biggl(\biggl({r\over Q}\biggr)^4\biggr).
\sectionit
$$
$$
f={Q^2(\sqrt{s}+\pi\sqrt{2})^2\over 2(1+\hbox{cosh}\pi)}+{\cal O}(r^2).
\sectionit
$$
Note that (4.22), (4.23) and (4.24) are exact for any $s$.

Let us consider the electric field obtained from (3.11). We have
$$
f^2+r^4=\lambda_0\hbox{exp}(-2\nu)(1+s^2)(\hbox{cosh}\psi_a
-\hbox{cos}\psi_b)^{-2},
\sectionit
$$
and
$$
\alpha\gamma={(\gamma^\prime)^2(f^2+r^4)\over 4[m^2+(\gamma-1)Q^2]}.
\sectionit
$$
This leads to the result:
$$
E_r={Q\gamma^\prime\over 2(m^2+(\gamma-1)Q^2)^{1/2}}.
\sectionit
$$
Using (3.4) and (3.6), we can calculate the invariant quantity:
$$
\Phi=F^{\mu\nu}F_{\mu\nu}.
$$
We find that
$$
\Phi=-{2Q^2\over r^4+f^2},
\sectionit
$$
which is finite at $r=0$. Moreover, if we define the dual tensor:
$\,^*F^{\mu\nu}=\epsilon^{\mu\nu\alpha\beta}F_{\alpha\beta}$, then it
follows that the other electromagnetic invariant,$\,^*F^{\mu\nu}F_{\mu\nu}$
equals zero. The electric field invariant $\Phi$ is finite in the presence of
the nonsymmetric gravitational
field. In fact, when any gauge field is incorporated into NGT, the resulting
combined field theory invariants will be made finite at $r=0$, because
of the geometrical properties of NGT at $r=0$.

It should be stressed at this stage that the {\it general solution} with
$\ell^2
\not=0$ is non-singular everywhere in spacetime. Our results
derived for $\ell^2=0$ can be extended to the general Vanstone or
Mann solution, depending in a non-analytic way on the dimensionless
parameter $s$.

For $Q=0$ and $s < 1$ the maximum red-shift is between $r=0$ and
$r=\infty$, and is given by
$$
z=\hbox{exp}\biggl({\pi\over 2\vert s\vert}+{s\over \vert s\vert}
+{\cal O}(s)\biggr)-1.
\sectionit
$$
A timelike Killing vector at spatial infinity remains timelike throughout
the spacetime, which means that our solutions are free of event horizons
and do not possess black holes.

The NGT curvature invariants such as the generalized Kretschmann scalar
are finite. This follows from the fact that the Mann, Vanstone and Wyman
solutions share the same form of expansion near $r=0$:
$$
\gamma= \gamma_0+\gamma_2r^2+...,
\sectionit
$$
$$
\alpha= \alpha_2r^2+\alpha_4r^4+...,
\sectionit
$$
$$
f=f_0+f_2r^2+... .
\sectionit
$$
Because the NGET solution takes this form a calculation shows that
all the curvature invariants are finite. For example, the generalized
Kretschmann invariant is given near $r=0$ by
$$
K=R^{\mu\nu\kappa\lambda}R_{\mu\nu\kappa\lambda}=
-{4\over f_0^4\alpha_2^4\gamma_0^4}\biggl[\alpha_2^4
f_0^2\gamma_0^4-\gamma_2^2\alpha_4^2\gamma_0^2f_0^4
-f_2^4\alpha_2^2\gamma_0^4
$$
$$
+4\alpha_2^3f_0f_2\gamma_0^4-\gamma_2^4\alpha_2^2f_0^4-
\alpha_2^2\gamma_0^4-
2\gamma_2^3\alpha_2\alpha_4\gamma_0f_0^4+6f_2^2\alpha_2^2\gamma_0^4\biggr]
+ {\cal O}(r^2).
\sectionit
$$
For the case $Q=0$, we find to leading order in $s$:
$$
K={\hbox{exp}(2\pi/s+4)\over 16m^4},\quad r=0,
\sectionit
$$
$$
={48m^2\over r^6},\quad r\rightarrow \infty.
\sectionit
$$

We note that the singularity caused by the vanishing of $\alpha(r)$ at
$r=0$ is a {\it coordinate} singularity, which can be removed by
transforming to another coordinate frame of reference. The curvature
invariants do not, of course, contain any coordinate singularities.

We can transform the standard line element:
$$
ds^2=\gamma dt^2-\alpha dr^2-r^2(d\theta^2+\hbox{sin}^2\theta d\phi^2)
\sectionit
$$
to a line element which is regular near $r=0$ by the transformation:
$$
\bar r={r^2\over m}.
\sectionit
$$
We obtain for $Q=0$:
$$
ds^2=\biggl[\gamma_0+{\cal O}\biggl({\bar r\over m}\biggr)\biggr]dt^2
-\biggl[{\gamma_0(1+{\cal O}(s^2))\over s^2}+{\cal O}
\biggl({\bar r\over m}\biggr)\biggr]d{\bar r}^2
- m\bar r(d\theta^2+\hbox{sin}^2\theta
d\phi^2).
\sectionit
$$
In this coordinate system, a radially directed photon near ${\bar r}=0$
has the finite coordinate velocity:
$$
{d{\bar r}\over dt}=s + {\cal O}(s^3).
\sectionit
$$

Let us now consider the proper volume obtained from (3.7):
$$
V_p=\int \sqrt{-g}drd\theta d\phi.
\sectionit
$$
By using (3.7), we get for $Q=\ell^2=0$ near $r=0$:
$$
\sqrt{-g}= {4\over 3}\hbox{exp}(-\pi/s - 2)r\hbox{sin}\theta,
\sectionit
$$
and
$$
V_p={16\pi r^2m\over s}\hbox{exp}(-\pi/s - 2).
\sectionit
$$
In comparison, EGT gives
$$
V_p={4\pi r^3\over 3}.
\sectionit
$$

The surface area and circumference of a body in the non-singular
NGT solution are given by $S=\hbox{Area}=4\pi r^2$ and $\hbox{Circumference}
=2\pi r$, respectively, which is
the same for the Schwarzschild solution in EGT.  The curvature invariants
are proportional to $(S/V_p)^2$ and they are finite constants at $r=0$.
The $V_p$ scales as the surface area for small $r$ which demonstrates
that a unique geometry exists at the origin. It is this unique geometry
which renders all fields finite at $r=0$. The proper volume near $r=0$
in NGT is infinitely larger than in EGT, and it is by this mechanism that
infinite energy is avoided.

The equations of motion of a test particle, in NGT , have been derived
from the matter response equations (2.19)$^{5}$:
$$
{du^\mu\over d\tau}+\left\{{\mu\atop \alpha\beta}\right\}u^\alpha u^\beta
={\ell^2_t\over m_t}{K^\mu}_\nu u^\nu+{e\over m_t}{F^\mu}_\nu u^\nu,
\sectionit
$$
where $u^\mu=dx^\mu/d\tau$, $\ell^2_t, m_t$ and $e$ are the test particle NGT
charge, mass and electric charge, respectively, and
$$
{K^\mu}_\nu= {1\over 2}\gamma^{(\mu\rho)}R_{[\rho\nu]}(\Gamma).
\sectionit
$$
Moreover,
$$
\left\{{\lambda\atop \alpha\beta}\right\}={1\over 2}\gamma^{(\lambda\rho)}
\biggl(g_{(\mu\rho),\nu}+g_{(\rho\nu),\mu}+g_{(\mu\nu),\rho}\biggr),
\sectionit
$$
and
$$
\gamma^{(\lambda\rho)}g_{(\lambda\sigma)}=\delta^\sigma_\nu.
\sectionit
$$

We have $R_{[10]}(\Gamma)=R_{[23]}(\Gamma)=0$ for $\ell^2=0$
and for $Q=0$, it follows from (4.41) that test particles follow geodesics
as in EGT:
$$
{du^\mu\over d\tau}+\left\{{\mu\atop \alpha\beta}\right\}u^\alpha u^\beta=0.
\sectionit
$$
For the non-singular solutions of NGT the spacetime is geodesically complete.

In EGT, it follows from the Hawking-Penrose theorem$^{20}$ that when
a star collapses it forms a trapped surface and this leads inevitably to
a singularity at $r=0$, provided the density $\rho$ and the pressure $p$
satisfy $\rho+3p > 0$. In NGT, the presence of repulsive forces for
$\vert s\vert > 0$
prevents the formation of singularities, trapped surfaces and event horizons,
while the positivity condition: $\rho+3p > 0$ is satisfied at all times.
Thus, even an infinitesimally small value of $s$ leads to a non-singular
theory of gravity free of black holes, which satisfies all know experimental
observation in gravity. It would be tempting to say that nature would
prefer such a rational description of spacetime compared to EGT, in which
singularities occur and observable quantities like densities and red-shifts
become infinite at points in spacetime.

Let us calculate the energy density of a charged body between $r=0$ and
$r=\infty$. We have
$$
{T^0}_0= {Q^2\over 8\pi(r^4+f^2)}.
\sectionit
$$
Integrating this over a volume in spherical polar coordinates gives
$$
{\cal E}=\int \sqrt{-g}{T^0}_0 dr d\theta d\phi={Q^2\over 2}\int^{\infty}_0
{(\alpha\gamma)^{1/2}\over (r^4+f^2)^{1/2}}dr
$$
$$
={Q^2\over 2}\int^{\infty}_0{y^\prime dr\over (4yQ^2+\lambda_0)^{1/2}}
={1\over 4}\biggl[(4Q^2+\lambda_0)^{1/2}-
(4\gamma_0Q^2+\lambda_0)^{1/2}\biggr].
\sectionit
$$
For the case when $\lambda_0=4(m^2-Q^2)$, we find that
$$
{\cal E}={1\over 2}\biggl\{m-\biggl[m^2-Q^2(1-\gamma_0)\biggr]^{1/2}\biggr\}.
\sectionit
$$
For $m >> Q$, this becomes
$$
{\cal E}={Q^2\over 2m}(1-\gamma_0)+{\cal O}
\biggl(\biggl({Q\over m}\biggr)^2\biggr).
\sectionit
$$
For the extremal case $Q=m$, we get
$$
{\cal E}=Q\biggl[{\pi\over \sqrt{2}(\sqrt{2}\pi+\sqrt{s})}\biggr].
\sectionit
$$

We have obtained the remarkable result that the total electromagnetic field
energy for a spherically symmetric charged particle is finite. In the past,
it was suggested that electrons had a finite size in order to overcome the
problem of divergences in field theory. This proposal suffered from the problem
that the Coulomb repulsive force would blow the particle apart. Here the
finiteness of the particle energy is not achieved by giving it a finite size,
but by increasing the proper volume near $r=0$. A correct description of
an electron must be based on a quantum field theory, so
we cannot expect a classical NGT desciption of the electron to be realistic.
In natural units $Q >> m$ for the electron, whereas in our solution
$Q \leq m$.
\vskip 0.2 true in
\setsection\proclaim 5. {\bf Conclusions}
\par
\vskip 0.2 true in
We have analyzed solutions to the NGET and NGT field equations, which are
non-singular everywhere in spacetime. The curvature invariants are finite
and there are no event horizons, which means that there are no black hole
solutions in the theory. There will exist stable, super dense objects that
replace black holes, which have large red-shifts at their surfaces. These
SDO's will radiate thermal and non-thermal radiation; they will not radiate
Hawking radiation and there will not be an information loss problem as
with black holes. This has profound implications for our understanding
of quantum gravity, for there is no need for a fundamental change in
quantum mechanics to resolve the problem of predictability
associated with black holes and Hawking radiation in EGT, because spacetime
is no longer separated into two disconnected regions described by two
different Hilbert spaces.

The problem of gravitional collapse is presently being investigated$^{21}$,
together with the singularity problem in early Universe cosmology.
A preliminary analysis reveals that there will be a non-zero acceleration of
the
coordinate scale factor $R(r,t)$ as a collapsing star approaches $r=0$,
yielding a non-singular final state of collapse. Similar conclusions can be
drawn about the non-singular nature of the cosmological solution at
$t=0$.

It is clear that in the light of the success of constructing a consistent
non-singular
theory of gravity without black holes, in the form of non-singular NGET
and NGT, the whole notion that black holes exist in nature must be
critically reconsidered. It is, of course, difficult to observationally
distinguish a SDO from a black hole. The SDO is kept stable
by the repulsive skew symmetric forces which supplement the standard
matter pressures
when the SDO reaches extemely high densities.
In the case of active galactic
nuclei, that have been purported to be large black holes$^{22}$, the event
horizon at $r=2m$ would be deep within the interior of the galaxy. Only
an unambiguous detection of a null surface at $r=2m$ would clearly
settle the issue, but such a null surface is hidden from the view of the
observer. The same is true of black hole candidates such as
Cygnus X-1$^{23}$ for which the event horizon is hidden from view by
Newtonian-like accretion disks. The estimated mass of the unseen companion
of Cyg X-1 is $M_x\sim 10-16\,M_{\odot}$, which is too large to be a
neutron star. But the criterion used to identify the dark companion with
a black hole
is based on EGT and the choice of an equation of state for matter at or greater
than nuclear densities. Studies have shown that an equation of state
derived from alternative exotic types of matter, such as soliton
stars$^{24}$, or an equation of state based on effective field theories
of bulk nuclear matter$^{25}$, can lead to stable compact objects
with masses in excess of $10^6\,M_{\odot}$. At present, there is no
known unique signature that distinguishes such EGT objects from an SDO
in NGT.

It is imperative that new
attempts be made to experimentally settle the question of whether genuine
event horizons exist in nature, because
an unambiguous answer will have profound implications for our understanding
of the nature of the geometry of spacetime.
\vskip 0.2 true in
{\bf Acknowledgements}
\vskip 0.2 true in
We thank M. Clayton, P. Savaria and G. Starkmann for helpful and stimulating
discussions. We thank the Natural Sciences and Engineering Research Council of
Canada for the support of this work.
\vskip 0.3 true in
\centerline{\bf References}
\vskip 0.2 true in
\item{1.}{A. Einstein, Ann. der Physik, {\bf 49},769 (1916); reprinted in:
H. A. Lorentz, A. Einstein, H. Minkowski, H. Weyl, The Principle
of Relativity, translated by W. Perrett, G. B. Jeffery, Dover Publications,
New York, p.164, 1923.}
\item{2.}{Albert Einstein: Philosopher--Scientist, edited by P. A. Schilpp,
Tudor Publishing Company, New York, p. 95, 1957.}
\item{3.}{A. Einstein, The Meaning of Relativity, 5th edition, Menthuen and
Company, London, p.127, 1951.}
\item{4.}{J. W. Moffat, Phys. Rev. D {\bf 19}, 3554 (1979); D {\bf 19},
3562 (1979); J. Math. Phys. {\bf 21}, 1978 (1980); Found. Phys.
{\bf 14}, 1217 (1984).}
\item{5.}{J. W. Moffat, Phys. Rev. D {\bf 35}, 3733 (1987);
J. W. Moffat and E. Woolgar, Phys. Rev. D{\bf 37}, 918 (1988).}
\item{6.}{For a review of NGT, see: J. W. Moffat, Proceedings of the
Banff Summer Institute on Gravitation, Banff, Alberta, August 12-25, 1990,
edited by R. Mann and P. Wesson, World Scientific, p.523, 1991.}
\item{7.}{K. Lake, Phys. Rev. Lett. {\bf 63}, 3129 (1992); B. Waugh and
K. Lake, Phys. Rev. D{\bf 38}, 4, 1315 (1988); P. S. Joshi and I. H. Dwivedi,
Commun. Math. Phys. {\bf 146}, 333 (1992); Lett. Math. Phys. {\bf 27},
235 (1993); I. H. Dwivedi and P. S. Joshi (to appear in Commun. Math.
Physics, 1994); D. M., Eardley in ``Gravitation in Astrophysics'', Plenum
Publishing Corporation, edited by B. Carter and J. B. Hartle, p.223, 1987.}
\item{8.}{S. Hawking, Phys. Rev. D{\bf 14}, 2460 (1976); Commun. Math.
Phys. {\bf 87}, 395 (1982).}
\item{9.}{J. W. Moffat, Procedings of the XXVIIIth Rencontre de Moriond,
Villars sur Ollon, Switzerland, January 30--February 6, 1993, edited by
J. Tran Thanh Van, T. Damour, E. Hinds and J Wilkerson, Editions Frontieres,
France, p.533, 1993.}
\item{10.}{N. J. Cornish and J. W. Moffat, University of Toronto preprint,
UTPT-94-04, 1994.}
\item{11.}{N. J. Cornish, University of Toronto preprint, UTPT-94-10, 1994.}
\item{12.}{L. Campbell, J. W. Moffat, and P. Savaria, Astrophys. J. {\bf 372},
241 (1991).}
\item{13.}{M. W. Kalinowski and G. Kunstatter, J. Math. Phys. {\bf 25}, 117
(1984); G. Kunstatter, J. Math. Phys. {\bf 25}, 2691 (1984).}
\item{14.}{R. B. Mann, J. Math. Phys. {\bf 26}, 2308 (1985).}
\item{15.}{A. Papapetrou, Proc. Roy. Ir. Acad. Sec. A {\bf 52}, 69 (1948).}
\item{16.}{M. Wyman, Can. J. Math. {\bf 2}, 427 (1950).}
\item{17.}{W. B. Bonnor, Proc. R. Soc. London {\bf A209}, 353 (1951);
{\bf A210}, 427 (1952).}
\item{18.}{J. R. Vanstone, Can. J. Math. {\bf 14}, 568 (1962).}
\item{19.}{H. Reissner, Ann. Phys. (Leipzig) {\bf 50}, 106 (1916);
G. Nordstrom, K. Ned. Akad. Wet. Versl. Gewone Vergad. Afd. Natuurkd.
{\bf 20}, 1238 (1918).}
\item{20}{R. Penrose, Phys. Rev. Lett. {\bf 14}, 57 (1965); S. Hawking,
Proc. R. Soc. London {A300}, 187 (1967); S. Hawking and G. Ellis,
Large Scale Structure of Space-Time, Cambridge University Press,
1973.}
\item{21.}{N. J. Cornish and J. W. Moffat (in preparation).}
\item{22.}{J. Kormendy, Astrophys. J. {\bf 325}, 128 (1988);
D. Richstone, G. Bower, and A. Dressler, Astrophys. J. {\bf 353}, 118 (1990).}
\item{23.}{A. P. Cowley, Annu. Rev. Astron. Astrophys. {\bf 30}, 287
(1992).}
\item{24.}{T. D. Lee, Phys. Rev. D{\bf 35}, 3637 (1987); T. D. Lee and
Y. Pang, {\it ibid.} D{\bf 35}, 3678 (1987).}
\item{25.}{S. Bahcall, B. W. Lynn, and S. B. Selipsky, Astrophys. J.
{\bf 362}, 251 (1990).}
\end